# Nonlocal and Quantum Tunneling Contributions to Harmonic Generation in Nanostructures: Electron Cloud Screening Effects

Michael Scalora[1], Maria Antonietta Vincenti[2], Domenico de Ceglia[2], Joseph W. Haus[2,3]

[1] Charles M. Bowden Research Center, AMRDEC, RDECOM, Redstone Arsenal, AL 35898-5000
[2] National Research Council - AMRDEC, Charles M. Bowden Research Center, Redstone Arsenal, AL 35898
[3] Electro-Optics Program, University of Dayton, Dayton, OH 45469-2951

Our theoretical examination of second and third harmonic generation from metal-based nanostructures predicts that nonlocal and quantum tunneling phenomena can significantly exceed expectations based solely on local, classical electromagnetism. Mindful that the diameter of typical transition metal atoms is approximately 3Å, we adopt a theoretical model that treats nanometer-size features and/or sub-nanometer size gaps or spacers by taking into account: (i) the limits imposed by atomic size to fulfill the requirements of continuum electrodynamics; (ii) spillage of the nearly-free electron cloud into the surrounding vacuum; and (iii) the increased probability of quantum tunneling as objects are placed in close proximity. Our approach also includes the treatment of bound charges, which adds crucial, dynamical components to the dielectric constant that are neglected in the conventional hydrodynamic model, especially in the visible and UV ranges, where interband transitions are important. The model attempts to inject into the classical electrodynamic picture a simple, perhaps more realistic description of the metal surface by incorporating a thin patina of free-electrons that screens an internal, polarizable medium.

## I. INTRODUCTION

It is generally recognized that theoretical studies of typical optical phenomena that take place at nanometer and sub-nanometer scales necessitate the adoption of methods that go beyond the usual approaches associated with classical electromagnetism. Two relevant examples are nonlocal effects and quantum tunneling phenomena. Plasmonic phenomena can occur between metallic objects and cavity walls that are in such close proximity that the electronic clouds nearly touch, and an applied electromagnetic field can induce electrons to tunnel between metal objects. Quantum tunneling phenomena have been addressed using numerically intensive, time-dependent density functional theory (TDDFT), to explore the limitations of classical theory [1-5]. The TDDFT has also been modified into a simpler method referred to as the Quantum Correction Model, which assigns to the gap region the same free-electron properties as the interacting metal components [2]. More recently, in this regard we have developed a Quantum Conductivity Theory, or QCT [6-8], that predicts linear and nonlinear, quantum-induced current densities in the gap region, either a vacuum or a dielectric material, such that the gap itself acquires additional linear and nonlinear optical properties.

While the induced quantum currents tend to limit field enhancement as a result of induced linear and nonlinear absorption, electron tunneling may also facilitate harmonic generation at rates that far exceed typical conversion efficiencies expected for metal nanostructures if quantum tunneling were neglected [9]. For practical purposes, the TDDFT is limited by the number of electronic wave functions that may be used to describe a nanostructure, and so the system under consideration must be small and made of the same metal [1-5, 10]. In contrast, the QCT [6-8] generally yields results similar to the TDDFT theory, uses no free parameters, and may be easily combined with Maxwell's equations to explore a wide variety of complex plasmonic systems composed of different metals and insulators, as well as nonlinear optical phenomena that arise as a consequence of quantum tunneling [9].

In addition to quantum tunneling, abrupt changes to the charge density at or near the surface can trigger nonlocal effects. These effects may be studied in a purely classical environment by relating the charge density to the pressure density of an ideal electron gas [11]. The assumption that the electron gas has a quantum nature yields a two-component plasma medium whose linear contribution coincides with the classical, ideal gas expression [12, 13], and with purely quantum mechanical contributions mostly to harmonic generation if the pump remains undepleted, and to additional nonlocal contributions if the pump energy is drained by a nonlinear conversion process. The result is that the dielectric constant turns into a function of frequency and wave vector, i.e. $\varepsilon = \varepsilon(\omega, \mathbf{k})$, and the polarization becomes a function of the field and its spatial derivatives. The effect "softens" the metal surface and smears charges and fields just beneath it. Local (i.e. $\varepsilon = \varepsilon(\omega)$), classical models predict an ever-increasing local field enhancement as the gap between metal components is reduced. In contrast, the inclusion of nonlocal effects in the hydrodynamic model [14-16] typically results in a reduction of the local field in the gap region, accompanied by field penetration that may be exploited to access the metal's nonlinearity [17].

Interest in the study of harmonic generation from metal surfaces has never abated since the early days of nonlinear optics [18-56]. The effective, second-order metal nonlinearity is usually decomposed as separate, tensorial surface and volume contributions [45-56] that have dipolar and quadrupolar origins. Our own, previous treatment of the problem [57, 58] was based on extending the hydrodynamic model [19, 31, 53], which treats conduction electrons only (e.g., 6s-shell for Au, 5s-shell for Ag), by including explicit, microscopic dipolar [57] and quadrupolar [58] contributions from bound electrons (5d-shell for Au and 4d-shell for Ag), and by making no a priori assumptions about what constitutes either surface or volume source. The classical Drude-Lorentz system that we use is well-behaved at the surface and yields no unmanageable singularities [51, 52]. The inclusion of the linear and nonlinear dynamics of bound electrons may certainly be viewed as an improvement to the hydrodynamic model, especially in the visible and UV ranges, where the dielectric function deviates significantly from Drude-like behavior. However, an additional, persistent issue is the size of typical surface features and gap or spacer thicknesses, which with modern atomic layer deposition techniques can easily approach and even be smaller than 1nm: one has to contend with atomic diameters (or lattice constants) that are of order 3Å, and an outer-shell, electron cloud that may extend several Å outside the last atomic, surface layer.

Like all classical models, the model exemplified in references [57-58] *does not* contemplate length scales or roughness on the order of the atomic thickness, and *should not* be expected to compare well with purely quantum mechanical approaches [10]. However, the situation may be mitigated by invoking an argument that permits the use of the classical, macroscopic equations in an atomic environment if sources are treated quantum mechanically [59, 60]. This approach may be summarized in Fig.1, where we illustrate scale drawings of a typical transition metal atom (inset at the top of the figure), complete with inner *d*-shells and outer *s*-orbital. A full-fledged, quantum mechanical Hartree-Fock approach that includes electron-electron interactions and screening may be used to calculate the wave functions associated with each orbital. The wave functions may then be used to deduce orbital radii [61, 62]. By orbital radius one generally refers to the distance from the nucleus to the maximum of the wave function (or most likely electron position), which may in fact have several nodes and be somewhat extended in space [63]. For example, the radius of the uppermost, 5*d*-orbital of gold is approximately $r_d$ =0.64Å, while the wave function of the 6*s*-shell peaks at $r_s$ =1.56Å [61, 62], or approximately one atomic radius. Cu and Ag display similar values (see caption of Fig.1). Then, for atoms arranged in a lattice, the simplest, most rudimentary picture that emerges is similar to the illustration at the bottom of Fig.1: nearly-free, outer *s*-shell electrons and bound (inner core) electrons permeate the entire volume, while all rows of atoms near the surface are slightly submerged under *s*-shell (conduction) electrons (tiny dots in the illustration) that spill outside the metal surface, and in so doing screen the internal medium.

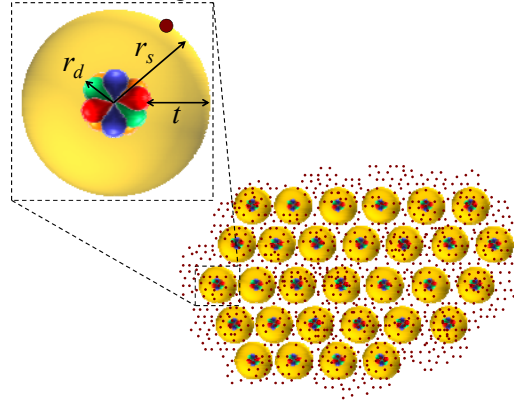

FIG. 1. (Color online) Top Left: Scale illustration of *d*- and *s*-orbitals within a single metal atom obtained via Hartree-Fock theory. The radii of the *d*- and *s*-orbitals, $r_d$ and $r_s$, respectively, correspond to the maxima of each calculated electronic wave function. For Au: $r_d$ =0.64Å and $r_s$ =1.56Å; Ag: $r_d$ =0.55Å and $r_s$ =1.53Å; Cu: $r_d$ =0.33Å and $r_s$ =1.37Å [61, 62], and $t$~1Å. Bottom Right: The electron cloud composed of outer s-shell electrons permeates the entire volume. At the same time, free-electrons that belong to atoms near the surface spill outside and screen the hard, ionic background.

Quantum mechanical calculations that assume a uniform, smooth, generic ionic background in fact predict an average, free-electron spill-out distance of approximately 2Å, which may be understood as roughly the midpoint of a rapidly rarefying medium, i.e. a decaying, electronic wave function whose tail may actually reach somewhat deeper into the surrounding vacuum [59, 64-67].

The information contained in Fig.1 suggests that a classical, Drude-Lorentz oscillator model may be modified to incorporate the basic ideas. One may assume that the medium is composed of an internal, uniform, polarizable mixture of free (Drude) and bound (Lorentz) electrons that extends as far as the outermost reaches of surface atoms' *d*-orbitals, and by a thin layer of *s*-shell, free electrons that screens the internal medium. Of course, this is a simplified view that seeks to combine the quantum properties of the atom with macroscopic field equations that are derived in a context where atomic-size or roughness must be averaged out, leaving behind only smooth surfaces. In a classical sense, the immediate consequence of the adoption of this physical picture means that the generic, metallic medium that we envision is characterized by a surface layer of finite thickness that has two boundaries: an internal surface, where the (linear and nonlinear) effects of bound charges are extinguished, and an outer surface grazed only by free electrons. For simplicity we assume the density of the outer, free-electron layer remains constant, although in reality a density gradient is to be expected. Density variations of the electron cloud as a function of distance from the hard surface can be easily included in the model. However, we expect the qualitative aspects of the problem to remain unchanged.

## II. BRIEF OUTLINE OF THE MODEL

The model that we use is based on a microscopic portrayal that begins with a collection of classical Drude-Lorentz oscillators that describe free and bound electrons coupled by the fields. While free charges can move about the entire volume, the motion of bound charges takes place around an equilibrium position that we identify as the radius of the *d*-orbital, effectively creating two surfaces. The equations of motion that we use are derived and described in details in references [57] and [58], and so here we provide broad motivation for the approach. In the absence of quantum tunneling, the generated second harmonic signal is triggered by free and bound charges alike, because both types of charges interact with the applied fields via intrinsically nonlinear Coulomb (electric) and Lorentz (magnetic) forces. We note that the nonlinear dynamics of bound charges is usually neglected. Free charges are also under the action of nonlinear convective and electron gas pressure forces. In addition, the model allows for multi-polar, nonlinear source distributions [58] as a result of slight distortions of the inner-core electron cloud resulting from electron screening. In contrast, while a small fraction of the third-harmonic signal always arises from a weak, cascaded process [57] ($3\omega=\omega+2\omega$), most of it originates from a bulk, third-order nonlinearity attributable to anharmonicities in the motion of bound charges [68].

The effects of the QCT theory are described in details in Refs.[6-8]. The theory suggests that the gap that separates metal objects fills with induced, linear and nonlinear currents that turn the vacuum or dielectric spacer into an effective medium that displays its own peculiar, linear and nonlinear optical properties. For instance, a vacuum gap approximately $g=0.8$nm thick displays an effective $\chi^{(2)} \sim i0.1$pm/V for adjacent objects composed of dissimilar metals like Au and Ag, and an effective $\chi^{(3)} \sim i10^{-20}$m$^2$/V$^2$ for either similar or dissimilar metals, increasing exponentially for smaller gaps [8, 9]. Even though these values may appear to be relatively small, the intensity inside the gap may be amplified thousands of times compared to incident values, thus catalyzing efficient nonlinear optical processes that can far outweigh the intrinsic nonlinearities of the metal. Our approach thus places free and bound charges on the same footing, and adds crucial linear and nonlinear dynamical components to the dielectric constant that are neglected in the conventional hydrodynamic model. The equations of motion are integrated in the time domain using a split-step, fast Fourier transform method that propagates the fields, combined with a predictor-corrector method to integrate the material equations [69]. The two-dimensional spatial grid consists of 208x6000 lattice sites discretized in unit cells 1Å×1Å; the temporal step is $3\times10^{-19}$sec. Reflected and transmitted conversion efficiencies are calculated by sampling the fields at the grid's edges (i.e. far field), and by normalizing the outgoing energy with respect to the total, incident pump energy.

As a final note, we point out that the model outlined above clearly attempts to account for atomic structure and size, electron spill-out from the surface, and quantum tunneling in order to paint a somewhat more realistic picture of physical phenomena that take place near the metal surface. However, the same may not be said of alternative approaches that introduce artifacts to treat the metal nonlocality [70]. For example, in reference [70] the nonlocal metal is replaced with a composite material made of local metal covered by a dielectric layer approximately 1Å thick, (i.e. the approximate Thomas-Fermi length, or in our case, $r_s - r_d$) that even includes the possibility of unphysical gain, ostensibly for the sole reason to ease the computational burden that the metal nonlocality imposes on complex geometrical arrangements. While this may be a clever way to solve a linear problem, the method also injects an arbitrary artifice that completely upsets the linear and nonlinear postures of surface currents and sources, including nonlinear, nonlocal contributions [53, 57]. Then, depending on its precise composition and thickness, the top layer may interfere and perhaps even negate quantum tunneling effects. Indeed, the modifications that we advocate are easily implemented and pivotal for nonlinear processes like harmonic conversion efficiencies, which can change drastically depending on surface properties.

## III. EXAMPLE CALCULATION OF NANOWIRE ARRAY

In Fig. 2 we depict two separate arrays of infinitely long, metal nanowires. Each nanowire is 10nm in radius (approximately 30 atomic diameters). For both arrays, adjacent cylinders are separated by a distance of $g=0.8$nm, and may be thought of as being composed of either a single metal or dissimilar metals. In Fig. 2(a) we show the way metals are normally treated: free and bound electrons are allowed to be present everywhere, so that *d*- and *s*-orbitals belonging to surface atoms overlap, i.e. $r_b = r_f$, where the subscripts *b* and *f* stand for bound and free, respectively. In Fig. 2(b) we show the alternative picture that emerges based on Fig. 1, which yields partially overlapping *d*- and *s*-orbitals belonging to surface atoms, so that $r_b < r_f$. The calculated linear transmission, reflection, and absorption spectra for the two scenarios in Fig. 2 are plotted in Fig. 3 for an Au-Ag array, in the local approximation. Nonlocal effects originate in the free-electron gas pressure contribution, lead to decreased local fields, and cause a generic blue-shift of the plasmonic band structure, with few additional qualitative or quantitative differences [9], at least in this geometrical arrangement. Palik's gold data [71] are first fitted in the range indicated in the figure using one Drude and one Lorentz oscillator, and are used to calculate the linear spectra of reflection, transmission and absorption for the array in Fig. 2(a) (solid curves in Fig. 3). The Lorentz component is then removed in a limited region to account for the free-electron-only green shell shown in Fig. 2(b) and linear spectra recalculated (dashed curves in Fig. 3). The results in Fig. 3 thus show that the linear optical properties of the two arrays displayed in Fig. 2 are practically indistinguishable. However, the practical impossibility to distinguish between linear behaviors, e.g. in reference [70], where an artificial, active-gain shell is introduced to describe the nonlocality, does nevertheless lead to large discrepancies in nonlinear optical properties that are entirely attributable to the slight geometrical differences between Figs. 2(a) and (2b). For simplicity, in what follows we will assume an Au-Ag grating to ultimately excite simultaneously both second- and third-order nonlinearities inside the gap [8, 9], and that the average thickness of the screening, free-electron layer is 2Å [59, 64-67]. We will compare the results for both types of arrays in order to assess the relevance of the free-electron buffer layer in nanostructures with surface features/gap sizes that approach atomic size, and how nonlocality manifests itself in the two cases.

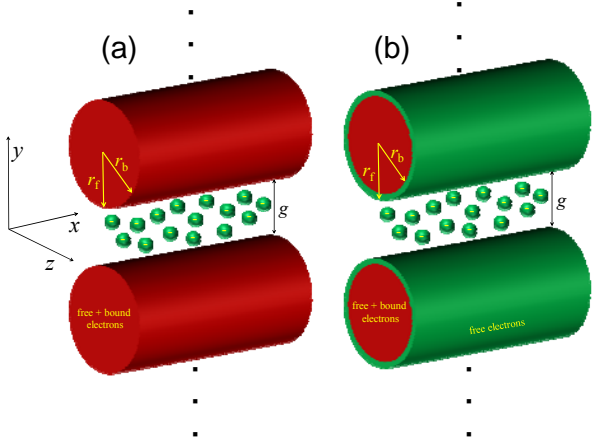

Fig.2. (Color online) (a) Metal array composed of a mixture of free and bound charges. *d*- and *s*-orbitals overlap, so that bound electron orbits graze the surface. (b) The illustration of Fig.1 yields a picture where an outer, free-electron shell approximately 2Å thick that covers each nanowire. The green particles inside the gap region of width *g* represent tunneling electrons.

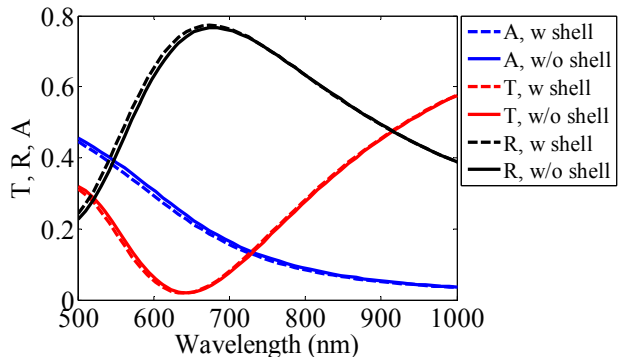

Fig.3. (Color online) Solid Curves: Reflection, Transmission, and Absorption vs. wavelength for the gold nanowire array in Fig. 2(a), where free and bound charges extend all the way to the surface. Dashed Curves: Reflection, Transmission, and Absorption for the gold nanowire array in Fig. 2(b), where bound charges are covered by a shallow, free electron layer.

In Table 1 we show the results for the predicted second harmonic generation (SHG) and third harmonic generation (THG) conversion efficiencies *without* quantum tunneling effects, with and without the free-electron buffer layer, with and without nonlocal effects. We assume that both metals exhibit an isotropic, third-order nonlinear response $\chi^{(3)}=10^{-18}$ $(m/V)^2$ [68]. Pump pulses are 25fs in duration, are tuned at 700nm, and have peak power of 1.5GW/cm$^2$. The second harmonic (SH) signal is tuned at 350nm and the

| **SH and TH conversion efficiencies** Peak Pump Intensity $I_\omega$ = 1.5 GW/cm$^2$ | | | |
|---|---|---|---|
| $\eta_{2\omega}$ | Local | $1.3\times10^{-8}$ | $1.8\times10^{-9}$ |
| | Nonlocal | $5\times10^{-9}$ | $2.2\times10^{-12}$ |
| $\eta_{3\omega}$ | Local | $2.2\times10^{-8}$ | $2\times10^{-8}$ |
| | Nonlocal | $4.6\times10^{-9}$ | $5\times10^{-9}$ |

third harmonic (TH) signal is tuned at 233nm. In general, SHG is far more sensitive than THG to surface phenomena because it depends intimately on the evolution and disposition of surface sources, given the centrosymmetric nature of the metal. Without the free-electron screening layer the nonlocal term smears charges and fields away from the surface just enough to reduce the magnitude of the field derivatives (Fig. 4), and hence the amplitudes of nonlinear surface sources, causing a reduction in conversion efficiencies by approximately a factor of two. In contrast, the introduction of nonlocality when a free-electron buffer layer is present makes the surface more elastic, voids all surface contributions due to bound charges [58], and reduces conversion efficiencies by nearly three orders of magnitude compared to its local counterpart. If we then compare only *nonlocal* predictions for SHG we find that the free-electron buffer layer suppresses surface contributions from bound charges very effectively and reduces conversion efficiencies by three orders of magnitudes. Most of the reduction of SHG conversion efficiency is due to the restriction of bound charges to the inner metal surface: the transition from the inner, red region to the green shell shown in Fig. 2(b) is much smoother compared to the vacuum/metal transition of Fig. 2(a), which reduces dramatically the influence of nonlinear, bound quadrupolar sources. Our calculations show that most of the reduction in SH efficiency is due to the absence of explicit, bound quadrupolar terms. This should not come as a surprise, since in the wavelength range of interest the dielectric constant is dominated by interband transitions, i.e. the bound electron cloud.

In contrast to SHG, the TH signal *in this particular case* is far less sensitive to the presence of the free-electron buffer layer because the transverse field component couples to the internal, bulk nonlinearity in nearly equal measures in both geometries. Just outside the nanowire, not only is the transverse electric field intensity nearly three orders of magnitude larger than the longitudinally polarized field, it is also shielded far less efficiently. In this case, nonlocal effects reduce conversion efficiencies by nearly a factor of three compared to the local case, because the nonlocality reduces overall field amplitudes.

The above observations on THG do not constitute general predictions because slight geometrical changes can strongly influence the outcome. For example, in reference [17] a gold nanowire of square cross section is placed approximately 1nm above a silver substrate. That arrangement strongly favors the longitudinal component of the field inside the gap region, triggering a localized surface plasmon with an evanescent tail that propels the field into the metal. In that environment, nonlocal effects can either: (i) increase THG by nearly three orders of magnitude, if the nanowire has no free-electron buffer layer, as in Fig. 2(a); or (ii) have no influence at all if a free-electron buffer layer only 1Å thick surrounds the nanowire, which is the approximate spatial separation between *d*- and *s*-shell electron orbits, and is sufficient to nearly completely suppress the enhancement of the field normal to the surface [17]. These considerations should serve as further cautionary notes that: (i) geometrical considerations always play an important role, and generalization should be avoided; and (ii) the presence of a shielding, free-electron outer layer can dramatically alter predicted, SHG and THG conversion efficiencies.

We now illustrate nonlocal effects on charge distribution. In Fig. 4 we show snapshots of the instantaneous, differential free charge density derivable from the continuity equation, in the local and nonlocal approximations, defined as $\delta n = \frac{n(\mathbf{r},t) - n_0}{n_0} = -\frac{1}{en_0}\nabla \cdot \mathbf{P}_f(\mathbf{r},t)$. Although this expression is valid whether or not the nanowires are surrounded by a thin, free-electron-only shell, here we treat the case illustrated in Fig. 2(b). $\delta n$ is normalized by the equilibrium (no applied field) charge density, $n_0$, $e$ is the electron charge,

$$\mathbf{P}_f(\mathbf{r},t) = \mathbf{P}_\omega(\mathbf{r},t)e^{-i\omega t} + \mathbf{P}_{2\omega}(\mathbf{r},t)e^{-2i\omega t} + \mathbf{P}_{3\omega}(\mathbf{r},t)e^{-3i\omega t} + c.c.$$

is the total polarization associated with free charges. If the pump remains undepleted, only the pump term contributes significantly. In Fig. 4(a) $\delta n$ is calculated in the local approximation; in Fig. 4(b) $\delta n$ is computed with the addition of the nonlocal electron gas pressure term. We note that in the nonlocal case the maximum amplitude of $\delta n$ is nearly four times smaller compared to its local counterpart, and slightly less confined to the surface. This smaller, local surface charge density value explains why the comparison of SHG conversion efficiencies between local and nonlocal cases in Fig. 4(b) strongly favors the local approximation: nonlinear surface and convective sources are proportional to $(\nabla \cdot \mathbf{P}_f)\mathbf{E}$ and $\dot{\mathbf{P}}_f(\nabla \cdot \mathbf{P}_f) + (\dot{\mathbf{P}}_f \cdot \nabla)\mathbf{P}_f$, respectively [54], and larger spatial derivatives lead to bigger SH conversion efficiencies.

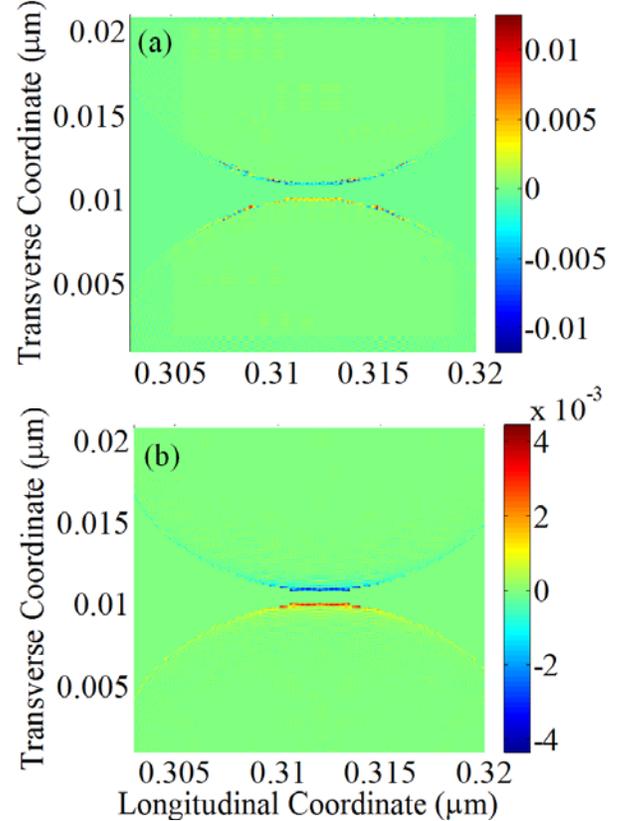

Fig. 4. (Color online) Differential, free charge density distribution $\delta n$ for: (a) local and (b) nonlocal cases. In (b) the metal surface is more elastic compared to (a), and can accommodate a smoother, but lower-amplitude charge distribution around the 2Å-thick nano-shell. Peak values of $\delta n$ are larger (by nearly a factor of four) and sharper in (a) compared to (b); this leads to larger SHG conversion efficiencies in the local case (a) compared to the nonlocal case (b). The distribution in (a) is numerically noiser compared to (b) because the field derivatives are very close to zero, giving rise to unphysical fluctuations inside the volume that have been averaged out.

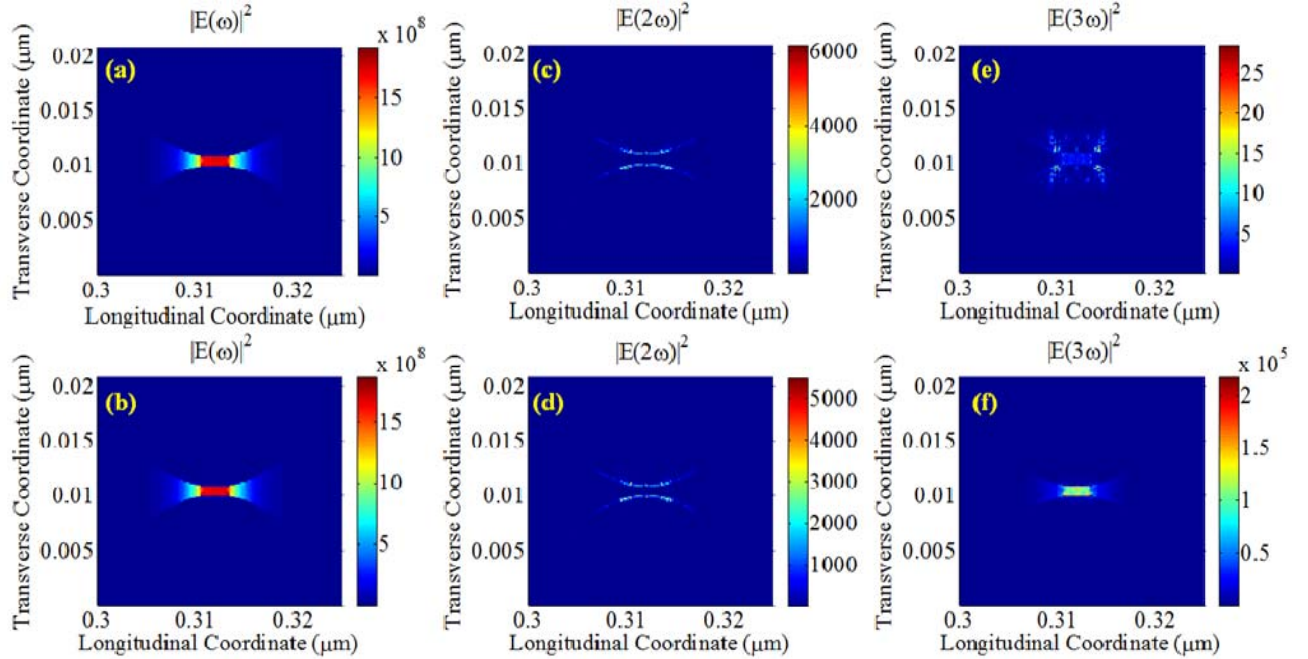

**Fig. 5.** (Color online) Electric field intensity distributions between two adjacent Au-Ag nanowires without (a, c, e) and with (b, d, f) quantum induced linear and nonlinear currents.

In Fig. 5 we show a snapshot of the spatial distribution of the electric field intensities when the peak of the incident pump pulse reaches the gold-silver grating with the free-electron shell in the nonlocal case, with and without quantum-induced currents. If adjacent nanowires are made of different metals, and the distance between nanowires is fixed at $g=0.8$nm, the currents inside the gap yield a linear dielectric constant $\varepsilon_\omega \approx 1 + i\,0.4$; $\chi^{(2)} \approx i\,0.1$ pm/V; and $\chi^{(3)}_\omega \sim \chi^{(3)}_{3\omega} \approx i10^{-20}$ (m/V)$^2$ [8-9]. We assume incident and generated fields are polarized along the array axis (as shown in Fig. 2). In the quantum tunneling case excitation of the gap region adds significantly to SHG and THG conversion efficiencies because the pump intensity becomes well-localized in the gap, with an enhancement factor of nearly three orders of magnitudes−Figs. 5(a) and (b). However, there are some peculiarities in the field localization properties that we now highlight. At moderate intensities (Fig. 5 corresponds to 0.4GW/cm$^2$ peak power), the peak pump field intensity is slightly reduced (less than 3%) by quantum-induced linear absorption. By the same token, in the quantum case the local TH field intensity inside the gap – Fig. 5(f) − is enhanced by nearly four orders of magnitude compared to the classical case – Fig. 5(e) −, a quantitative aspect that is also reflected in a corresponding increase in THG conversion efficiency, as reported in reference [9]. TH field localization inside the gap is also highly suggestive of the fact that for the most part nonlinear sources are distributed inside the gap, where $\chi^{(3)}_{3\omega} \approx i10^{-20}$ (m/V)$^2$, thus overwhelming THG arising from within the metal. However, the SH field localization properties are perhaps the most peculiar: indeed, quantum-induced currents increase conversion efficiency by nearly three orders of magnitudes [9], notwithstanding the fact that the local field intensity decreases by nearly 10% (compare amplitudes in Figs. 5(c) and (d)), with field localization properties that are practically unchanged relative to the absence of quantum tunneling. Put another way, a quantum gap is equivalent to a gap doped with a nonlinear material, thus comparable to the introduction of an effective dipolar contribution to the scattered SH light, forbidden in the classical representation. As a consequence, the presence of nonlinear quantum tunneling makes the structure a far more efficient radiator of SH light, even though field localization properties appear to change little compared to the classical case. We believe that this phenomenon is a unique marker of quantum tunneling. While linear effects of quantum tunneling are inherently subtle and hardly distinguishable from nonlocal effects, the nonlinear response is drastically altered by quantum tunneling, especially for SH light, whose nature is forcibly converted from quadrupolar to dipolar.

One needs to bear in mind that material dispersion, and thus incident wavelength, are important factors that determine the evolution of the harmonic fields. In general, in the wavelength range under consideration (below 700nm) the linear dielectric response of the metal at the harmonic wavelengths (deep in the UV range) is dominated by the dynamics of bound electrons. However, the nonlinear response in fact appears to be regulated by the presence of a screening, free-electron layer, which becomes the dominant feature in harmonic generation. As an example, pumping the grating at 600nm increases THG (at 200nm) by two orders of magnitude compared to pumping at 700nm, thanks to a combination of improved resonance conditions and field penetration depth, and reduced $\text{Im}(\varepsilon)$ for both Ag and Au.

Finally, we note that according to the QCT model [6-9] the magnitude of the quantum-induced coefficients increases at near-exponential rates for decreasing gap sizes. For example, according to the model a gap $g$=0.6nm wide will display increased linear absorption and nearly two orders of magnitude enhancement in the nonlinear coefficients compared to a 0.8nm gap. By the same token, a different geometrical arrangement may offer far improved field localization characteristics, as in reference [17], for example, where the local field intensity is enhanced more than $10^4$ times, which may suffice to trigger quantum tunneling events for slightly larger gap sizes [8].

## IV. CONCLUSIONS

We have presented a theoretical model that allows the study of linear and nonlinear optical phenomena like SHG and THG from nanoplasmonic environments in a context that takes into account: (i) linear and nonlinear dynamics of the bound electron cloud; (ii) electron spill-out effects and resultant screening of an internal, polarizable medium; and (iii) electronic quantum tunneling effects that induce linear and nonlinear currents between two metal objects placed in close proximity. We have investigated harmonic generation and compared the results in the local and nonlocal approximations, with and without electron screening conditions. In the absence of quantum tunneling, which for vacuum translates to gap sizes of order 1nm, and up to ~2nm for appropriate dielectric materials [8], our results suggests that both SHG and THG are sensitive to the geometry, the screening effects of a free-electron cloud that surrounds the internal medium, and nonlocal effects. For sufficiently small gap sizes, harmonic generation originating inside the quantum gap can easily overcome by several orders of magnitude the amplitudes of the harmonic signals arising from the intrinsic nonlinearities of the metal [9]. Our results thus suggests that both quadratic and cubic nonlinear effects may be triggered by quantum tunneling and may be observed in the far field at practical light irradiance levels (~1GW/cm$^2$), using femtosecond pulses, from metallic nanostructures with gap sizes of the order of 1nm.

## ACKNOWLEDGEMENTS

This research was performed while the authors J. W. Haus, M. A. Vincenti and D. de Ceglia held a National Research Council Research Associateship award at the U.S. Army Aviation and Missile Research Development and Engineering Center. We thank V. Roppo, M. Grande, C. Ciracì and D. Smith for fruitful discussions.